\documentclass{ifacconf}

\usepackage{graphicx}      % include this line if your document contains figures
\usepackage{natbib}        % required for bibliography
\usepackage{bm}
\usepackage{amsmath}
\usepackage{amssymb}
\usepackage{float}
\usepackage{hhline}
\usepackage{xcolor}
\usepackage{mathrsfs}
\usepackage{arydshln}
\usepackage{pdflscape}
\usepackage{tikz}
\usetikzlibrary{plotmarks,arrows,automata}
\newcommand{\tikzline}[1]{(\protect\tikz[baseline=-0.6ex,x=1pt,y=1pt]{ \protect\draw[#1] [-] (0,0) -- (10,0);})}
\newcommand{\tikzdashedline}[1]{(\protect\tikz[baseline=-0.6ex,x=0.9pt,y=1pt]{ \protect\draw[#1,dashed] [-] (0,0) -- (10,0);})}

\newcommand{\tikzsquare}[1]{ (\protect\tikz[baseline=0.2ex,x=1pt,y=1pt]{\protect\draw[fill=#1,draw=none,rounded corners= 1pt] (0,0) rectangle (5.5,5.5);})}

\newcommand{\QEDA}{\hfill\ensuremath{\blacksquare}}
\newcommand{\norm}[1]{\left\lVert#1\right\rVert}
\newcommand{\abs}[1]{\left\lvert#1\right\rvert}

\definecolor{MatlabRed}{rgb}     {0, 0, 0}
\definecolor{MatlabBlue}{rgb}    {0, 0, 0}
\definecolor{MatlabGreen}{rgb}   {0, 0, 0}
\definecolor{MatlabBlue2}{rgb}    {0     , 0.4470, 0.7410}
\definecolor{MatlabRed2}{rgb}     {0.8500, 0.3250, 0.0980}
\definecolor{MatlabGreen2}{rgb}   {0.4660, 0.6740, 0.1880}
\definecolor{MatlabYellow}{rgb}  {0.9290, 0.6940, 0.1250}
\definecolor{MatlabPurple}{rgb}  {0.4940, 0.1840, 0.5560}
\definecolor{MatlabBabyBlue}{rgb}{0.3010, 0.7450, 0.9330}
\definecolor{MatlabGray}{rgb}{0.5, 0.5, 0.5}
\definecolor{MatlabLightGray}{rgb}{0.75, 0.75, 0.75}
\definecolor{MatlabBlack}{rgb}{0, 0, 0}
\definecolor{MatlabLightGray4}{rgb}{0.875, 0.875, 0.875}
\definecolor{MatlabLightGray3}{rgb}{0.85, 0.85, 0.85}
\definecolor{MatlabLightGray2}{rgb}{0.775, 0.775, 0.775}
\definecolor{MatlabLightGray1}{rgb}{0.7, 0.7, 0.7}
\definecolor{MatlabGray20}{rgb}{0.2, 0.2, 0.2}
\definecolor{MatlabGray30}{rgb}{0.3, 0.3, 0.3}
\definecolor{MatlabGray40}{rgb}{0.4, 0.4, 0.4}
\definecolor{MatlabGray50}{rgb}{0.5, 0.5, 0.5}
\definecolor{MatlabGray60}{rgb}{0.6, 0.6, 0.6}
\definecolor{MatlabGray70}{rgb}{0.7, 0.7, 0.7}
\definecolor{MatlabGray80}{rgb}{0.8, 0.8, 0.8}
\definecolor{MatlabGray85}{rgb}{0.85, 0.85, 0.85}
\definecolor{MatlabGray90}{rgb}{0.9, 0.9, 0.9}

\newcounter{defi_count}
\newtheorem{defi}[defi_count]{Definition}
\theoremstyle{nonumberplain}
\newtheorem{proof-wo}{Proof}

\counterwithin*{section}{part}

\usepackage[capitalize]{cleveref}
\crefname{equation}{}{}
\Crefformat{figure}{#2Fig.~#1#3}
\Crefmultiformat{figure}{Figs.~#2#1#3}{ and~#2#1#3}{, #2#1#3}{ and~#2#1#3}
\crefrangelabelformat{figure}{#3#1#4--#5\crefstripprefix{#1}{#2}#6}
\Crefrangeformat{figure}{Figs.~#3#1#4--#5#2#6}

\usepackage{eso-pic}
\AddToShipoutPictureBG*{%
\AtPageUpperLeft{%
\setlength\unitlength{1in}%
\hspace*{\dimexpr0.5\paperwidth\relax}
\makebox(0,-1.75)[c]{
\begin{tabular}{c c}
Koen Classens \emph{et al.},
Direct Shaping of Minimum and Maximum Singular Values:\\ An $\mathcal{H}_{-}/\mathcal{H}_{\infty}$ Synthesis Approach for Fault Detection Filters, \\
To appear in
{\em Proceedings of the IFAC 22st Triennial World Congress}, Yokohama, Japan, 2023,
uploaded to arXiv \today \\
\end{tabular}}}}

\begin{document}
\begin{frontmatter}

\title{Direct Shaping of Minimum and Maximum Singular Values: An $\mathcal{H}_{-}/\mathcal{H}_{\infty}$ Synthesis Approach for Fault Detection Filters}

%\thanks[footnoteinfo]{Sponsor and financial support acknowledgment
%goes here. Paper titles should be written in uppercase and lowercase
%letters, not all uppercase.}

\author[First,Second]{Koen Classens} 
\author[First]{W.P.M.H. (Maurice) Heemels}
\author[First,Third]{Tom Oomen}

\address[First]{Department of Mechanical Engineering, Eindhoven University of Technology, Eindhoven, The Netherlands}
\address[Third]{Delft Center for Systems and Control, Dept. 3mE, Delft University of Technology, The Netherlands}
\address[Second]{(e-mail: k.h.j.classens@tue.nl)}

\begin{abstract}                % Abstract of not more than 250 words.
The performance of fault detection filters relies on a high sensitivity to faults and a low sensitivity to disturbances. The aim of this paper is to develop an approach to directly shape these sensitivities, expressed in terms of minimum and maximum singular values. The developed method offers an alternative solution to the $\mathcal{H}_{-}/\mathcal{H}_{\infty}$ synthesis problem, building upon traditional multiobjective synthesis results. The result is an optimal filter synthesized via iterative convex optimization and the approach is particularly useful for fault diagnosis as illustrated by a numerical example.
\end{abstract}


\end{frontmatter}
%===============================================================================
\section{Introduction} \label{sec:INTRODUCTION} 
Fault detection and isolation (FDI) is highly important in many control applications which are becoming increasingly more demanding and more complex. In particular, FDI is important for the high-tech production industry which is shifting towards predictive maintenance strategies. This paradigm shift is motivated as a result of the high costs associated with unscheduled downtime. In this context, real-time fault diagnosis of complex closed-loop controlled multi-input multi-output (MIMO) systems is the foundation for effective targeted maintenance and optimal scheduling of downtime.

It is commonly recognized that satisfactory performance of model-based FDI systems is only achievable by a balanced trade-off. Techniques based on $\mathcal{H}_{\infty}$ optimization and $\mu$ design have been developed and applied, see, e.g.,  \cite{sadrniaRobustInftyMu1997,dingUnifiedApproachOptimization2000,zhongLMIApproachDesign2003}. However, these $\mathcal{H}_{\infty}$ approaches do not directly account for this trade-off as  $\mathcal{H}_{\infty}$ is only a measure for maximum gain. Apart from rejecting disturbance, noise and being insensitive to model uncertainties, the fault diagnosis system needs to be as sensitive to faults as possible. Hence, fault sensitivity needs to be addressed explicitly during design.

One way to enforce sensitivity to faults is to reformulate the problem, see \cite{henryTheoriesDesignAnalysis2021}, or by reformulating the problem as a fault estimation problem, see \cite{stoustrupFaultEstimationStandard2002}. In this way, the problem can still be solved as a fictitious $\mathcal{H}_{\infty}$ problem, however, undesired conservatism may be introduced. Alternatively, more direct $\mathcal{H}_{-} / \mathcal{H}_{\infty}$ approaches are attractive as the trade-off is explicitly embedded in the problem formulation. $\mathcal{H}_{-} / \mathcal{H}_{\infty}$ filter design can be pursued via factorization approaches as in \cite{dingUnifiedApproachOptimization2000}, or, e.g., via Riccati equations, see \cite{liuOptimalSolutionsMultiobjective2007}. In particular LMI formulations are of interest due to the relative ease to incorporate additional design objectives, see \cite{wangLMIApproachIndex2007,houLMIApproachHinfty1996}. In addition, LMI methods are well established for controller synthesis and observer design as demonstrated in \cite{schererMultiobjectiveOutputfeedbackControl1997,schererLinearMatrixInequalities2020}. However, these methods are not specifically tailored to FDI problems.

Although important progress has been made in fault detection for complex engineered systems, at present accurate FDI for complex systems is hampered by lack of compatible synthesis tools. The aim of this paper is to develop an alternative $\mathcal{H}_{-} / \mathcal{H}_{\infty}$ synthesis algorithm, building upon traditional multiobjective synthesis results originating from controller design. The theory builds upon the notion of minimum gain and allows to directly shape the minimum and maximum singular values of the complete system. Hence, the contribution of this paper is twofold: 1) Development of an alternative approach for FDI design, 2) Development of the associated synthesis problem.

This paper is organized as follows. The paper proceeds with notation and the required preliminaries, listed in \Cref{sec:PRELIMINARIES}. The underlying subproblem for multiobjective filter design is presented in a generic manner in \Cref{sec:GENERIC PROBLEM FORMULATION}. Subsequently, the design specifications for fault diagnosis filter design, the relevant matrix inequalities, and its synthesis procedure are described in \Cref{sec:DESIGN SPECIFICATIONS AND SYNTHESIS FOR FAULT DIAGNOSIS}. A numerical fault diagnosis case study, presented in \Cref{sec:NUMERICAL EXAMPLE}, illustrates the effectiveness of the proposed approach and finally, a conclusion is drawn in \Cref{sec:CONCLUSION}.
%%%%%%%%%%%%%%%%%%%%%%%%%%%%%%%%%%%%%%%%%%%%%%%%%%%%%%%%%%%%%%%%%%%%%%%%%%%%%%%%
\section{Preliminaries} \label{sec:PRELIMINARIES}
\subsection{Notation}
%\atodo{Nalopen of we al het onderstaande gebruiken}
Positive definiteness and positive semidefiniteness of a matrix $A$ are denoted by $A \succ 0$ and $A \succeq 0$, respectively. Similarly, $A \prec 0$ and $A \preceq 0$ denote negative definite and negative semidefinite matrices, respectively. The sets of all nonnegative and positive integers are denoted $\mathbb{N}$ and $\mathbb{N}_{\geq 0}$. The sets of real numbers and nonnegative real numbers are indicated by $\mathbb{R}$ and $\mathbb{R}_{\geq 0}$. The set of $n$ by $n$ symmetric matrices is denoted as $\mathbb{S}^{n}$. By $\norm{\cdot}$ the Euclidean norm is defined. Repeated blocks within symmetric matrices are replaced by $*$ for brevity and clarity. The identity matrix is written as $I$ and a matrix of zeros is written as $0$. The maximum and minimum singular values of the matrix $A$ are denoted by $\bar{\sigma}(A)$ and $\underline{\sigma}(A)$, respectively. The real rational subspace of $\mathcal{H}_{\infty}$ is denoted by $\mathcal{RH}_{\infty}$. $y \in \mathcal{L}_{2}$ if $\norm{y}_{2}^{2} = \int_{0}^{\infty} y^{\top}(t) y(t) \mathrm{d} t < \infty$. $y \in \mathcal{L}_{2e}$ if $\norm{y}_{2T}^{2} = \int_{0}^{T} y^{\top}(t) y(t) \mathrm{d} t < \infty$, $T \in \mathbb{R}_{\geq 0}$. %, where $y_{T}(t) = y(t)$ for $0 \leq t \leq T$ and $y_{T} = 0$ for $T<t$.

\subsection{Minimum and Maximum Gain}

%\begin{defi}
%\textbf{Minimum gain} (\cite{bridgemanMinimumGainLemma2015})
%A causal continuous-time linear time-invariant (LTI) system, $G: \mathcal{L}_{2e} \rightarrow \mathcal{L}_{2e}$, has minimum gain $\nu \geq 0$ if for all initial states $x_{0} \in \mathbb{R}^{n}$, and inputs $u \in \mathcal{L}_{2e}$,
%\begin{equation}
%\norm{Gu}_{2T} - \nu \norm{u}_{2T} \geq \beta.
%\end{equation}
%The least conservative minimum gain is the supremum of $0 \leq \nu < \infty$ with $\beta \in \mathbb{R}$.
%\end{defi}

\begin{lem} \label{lem:mingain}
\textbf{Minimum gain} (\cite{liuOptimalSolutionsMultiobjective2007,wangLMIApproachIndex2007}) %A causal continuous-time LTI system, represented by the minimum phase transfer matrix $G(s) \in \mathbb{C}^{n_{y} \times n_{u}}$, has a minimum gain $\nu$ if and only if
%\begin{equation}
%0 \leq \nu \leq \inf_{\omega \in \mathbb{R}_{\geq 0}} \underline{\sigma} (G(j \omega)).
%\end{equation}
The smallest gain of the continuous-time LTI system $G: \mathcal{L}_{2e} \rightarrow \mathcal{L}_{2e}$, that is the $\mathcal{H}_{-}$ index, is defined as
\begin{equation}
\norm{G}_{-} = \inf_{\omega \in \mathbb{R}_{\geq 0}} \underline{\sigma}(G(j \omega)).
\end{equation}
The minimum gain is not a norm and therefor named the $\mathcal{H}_{-}$ index.
\end{lem}

\begin{defi}
\textbf{Maximum gain} (\cite{skogestadMultivariableFeedbackControl2001})
The $\mathcal{H}_{\infty}$ norm of the continuous-time LTI system $G: \mathcal{L}_{2e} \rightarrow \mathcal{L}_{2e}$, denoted as $\norm{G}_{\infty}$, is given by
%\begin{equation}
%\norm{G}_{\infty} = \sup_{u \in \mathcal{L}_{2}, u \neq 0} \frac{\norm{G u}_{2}}{\norm{u}_{2}},
%\end{equation}
%or equivalently, 
\begin{equation}
\norm{G}_{\infty} = \sup_{\omega \in \mathbb{R}_{\geq 0}} \bar{\sigma} (G(j \omega)).
\end{equation}
\end{defi}

\begin{lem}
\textbf{Minimum Gain Lemma} (\cite{bridgemanMinimumGainLemma2015,caverlyOptimalOutputModification2018})
Consider a continuous-time LTI system $G: \mathcal{L}_{2e} \rightarrow \mathcal{L}_{2e}$, with state space realization $(\mathcal{A},\mathcal{B},\mathcal{C},\mathcal{D})$, where $\mathcal{A} \in \mathbb{R}^{n \times n}$, $\mathcal{B} \in \mathbb{R}^{n \times m}$, $\mathcal{C} \in \mathbb{R}^{p \times n}$, and $\mathcal{D} \in \mathbb{R}^{p \times m}$. The system $G$ has minimum gain $\nu$ under the following sufficient condition. There exists $X \in \mathbb{S}^{n}$ and $\nu \in \mathbb{R}_{\geq 0}$, where $X \succeq 0$, such that
\begin{equation}
\begin{bmatrix}
X\mathcal{A} + \mathcal{A}^{\top} X - \mathcal{C}^{\top} \mathcal{C} & X \mathcal{B} - \mathcal{C}^{\top} \mathcal{D} & 0 \\
* & - \mathcal{D}^{\top} \mathcal{D} & \nu I \\
* & * & - I
\end{bmatrix} \prec 0. \label{eq:BMI_mingain}
\end{equation}
\end{lem}

\begin{lem} \label{lem:BRL}
\textbf{Bounded Real Lemma} (\cite{gahinetLinearMatrixInequality1994})
Consider a continuous-time LTI system $G: \mathcal{L}_{2e} \rightarrow \mathcal{L}_{2e}$, with state space realization $(\mathcal{A},\mathcal{B},\mathcal{C},\mathcal{D})$, where $\mathcal{A} \in \mathbb{R}^{n \times n}$, $\mathcal{B} \in \mathbb{R}^{n \times m}$, $\mathcal{C} \in \mathbb{R}^{p \times n}$, and $\mathcal{D} \in \mathbb{R}^{p \times m}$. The inequality $\norm{G}_{\infty} < \gamma$ holds under the following necessary and sufficient conditions. There exists $X \in \mathbb{S}^{n}$ and $\gamma \in \mathbb{R}_{>0}$, where $X \succ 0$, such that
\begin{equation}
\begin{bmatrix}
X\mathcal{A} + \mathcal{A}^{\top} X & X \mathcal{B} & \mathcal{C}^{\top} \\
* & -\gamma^{2} I & \mathcal{D}^{\top } \\
* & * & - I
\end{bmatrix} \prec 0. \label{eq:BMI_maxgain}
\end{equation}
\end{lem}

%\begin{lem}
%\textbf{Minimum gain lemma} \cite{bridgemanMinimumGainLemma2014}
%Consider a continuous-time LTI system $G: \mathcal{L}_{2e} \rightarrow \mathcal{L}_{2e}$, with state space realization $(A,B,C,D)$, where $A \in \mathbb{R}^{n \times n}$, $B \in \mathbb{R}^{n \times m}$, $C \in \mathbb{R}^{p \times n}$, and $D \in \mathbb{R}^{p \times m}$. The system $G$ has minimum gain $\nu$ under the following sufficient condition. There exists $X \in \mathbb{S}^{n}$ and $\nu \in \mathbb{R}_{>0}$, where $X \succ 0$, such that
%\begin{equation}
%\begin{bmatrix}
%XA + A^{\top} X - C^{\top} C & X B - C^{\top} D & 0 \\
%* & - D^{\top} D & \nu I \\
%* & * & - I
%\end{bmatrix} < 0. \label{eq:BMI_mingain}
%\end{equation}
%\end{lem}

\section{$\mathcal{H}_{-}/\mathcal{H}_{\infty}$ Approach to Fault Detection} \label{sec:GENERIC PROBLEM FORMULATION}
%\acomment{Tom: A new approach to Hinf/H- optimization for FDI design}
This section introduces the generalized plant formulation and introduces the relevant notation for transformation of variables and objective channel selection. In addition, the underlying multiobjective synthesis subproblem is described.

\subsection{General Closed-loop Interconnection}
\begin{figure}[b]
\centering
\includegraphics[]{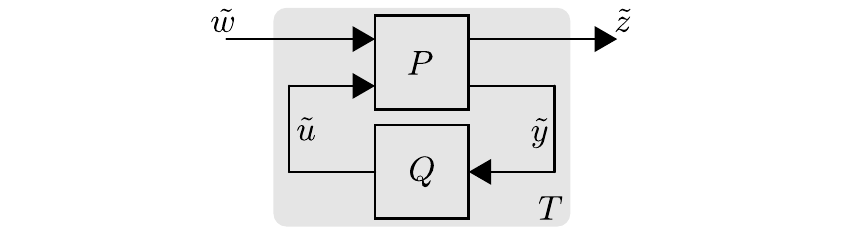}
%		\tikzexternaldisable   % Comment when tikzexternalize is off
\caption{Generalized plant $P: \begin{bmatrix}
\tilde{w}^{\top} & \tilde{u}^{\top}
\end{bmatrix}^{\top} \rightarrow \begin{bmatrix}
\tilde{z}^{\top} & \tilde{y}^{\top}
\end{bmatrix}^{\top}$ and filter $Q: \tilde{y} \rightarrow \tilde{u}$. The closed-loop system with performance channel $T: \tilde{w} \rightarrow \tilde{z}$ is highlighted in \tikzsquare{MatlabGray80}. }
% 		\tikzexternalenable    % Comment when tikzexternalize is off
\label{fig:CL_System}
\end{figure}

%\atodo{Consistent met of zonder tilde notatie. Notatie conflict met normale $u$, $y$ uit de std regellus.}
%\atodo{$Q$ is gedefinieerd met Ac, Bc, Cc, Dc, xc. Eventueel oplossen als tijd over.}
%\atodo{Check change of variables with $D_{22}$ t.o.v. Scherer}
Consider the generalized plant $P$, as depicted in \Cref{fig:CL_System},
with generalized disturbance channel $\tilde{w}$, generalized performance channel $\tilde{z}$, input $\tilde{u}$, and output $\tilde{y}$, which admits the state-space realization
\begin{equation}
\left[\begin{array}{c}
\dot{x} \\ \hline \tilde{z} \\ \tilde{y}
\end{array}\right] = \left[\begin{array}{c|cc}
A & B_{1} & B_{2} \\ \hline C_{1} & D_{11} & D_{12} \\ C_{2} & D_{21} & D_{22}
\end{array}\right] \left[\begin{array}{c}
x \\ \hline \tilde{w} \\ \tilde{u}
\end{array}\right]. \label{eq:Ptilde_eq}
\end{equation}
The to be designed filter $Q$ is any finite dimensional LTI system described as
\begin{equation}
\left[\begin{array}{c}
\dot{x}_{c} \\ \tilde{u}
\end{array}\right] = \left[\begin{array}{cc}
\textcolor{MatlabGreen}{A_{c}} & \textcolor{MatlabGreen}{B_{c}} \\ \textcolor{MatlabGreen}{C_{c}} & \textcolor{MatlabGreen}{D_{c}} 
\end{array}\right] \left[\begin{array}{c}
x_{c} \\ \tilde{y}
\end{array}\right].
\end{equation}
In particular, the state dimension of the filter $Q$, $x_{c}$, is not decided upon in advance. Let $T : \tilde{w} \rightarrow \tilde{z}$ denote the closed-loop transfer function, formed by the lower linear fractional transformation (LFT) between $P$ and $Q$. The closed-loop system $T = \mathcal{F}_{l} (P, Q)$, and admits the description
\begin{subequations}
\begin{align}
\dot{x}_{\mathrm{cl}} &= \textcolor{MatlabGreen}{\mathcal{A}} x_{\mathrm{cl}} + \textcolor{MatlabGreen}{\mathcal{B}} \tilde{w}, \\
\tilde{z} &= \textcolor{MatlabGreen}{\mathcal{C}} x_{\mathrm{cl}} + \textcolor{MatlabGreen}{\mathcal{D}} \tilde{w},
\end{align}
\end{subequations}
where
\begin{subequations}
\begin{align}
\textcolor{MatlabGreen}{\mathcal{A}} &= \begin{bmatrix}
A + B_{2} \textcolor{MatlabGreen}{D_{c}} \textcolor{MatlabGreen}{\bar{D}} C_{2} & B_{2} \left( I + \textcolor{MatlabGreen}{D_{c}} \textcolor{MatlabGreen}{\bar{D}} D_{22} \right) \textcolor{MatlabGreen}{C_{c}} \\
\textcolor{MatlabGreen}{B_{c}} \textcolor{MatlabGreen}{\bar{D}} C_{2} & \textcolor{MatlabGreen}{A_{c}} + \textcolor{MatlabGreen}{B_{c}} \textcolor{MatlabGreen}{\bar{D}} D_{22} \textcolor{MatlabGreen}{C_{c}}
\end{bmatrix}, \\
\textcolor{MatlabGreen}{\mathcal{B}} &= \begin{bmatrix}
B_{1} + B_{2} \textcolor{MatlabGreen}{D_{c}} \textcolor{MatlabGreen}{\bar{D}} D_{21} \\ \textcolor{MatlabGreen}{B_{c}} \textcolor{MatlabGreen}{\bar{D}} D_{21} 
\end{bmatrix}, \\
\textcolor{MatlabGreen}{\mathcal{C}} &= \begin{bmatrix}
C_{1} + D_{12} \textcolor{MatlabGreen}{D_{c}} \textcolor{MatlabGreen}{\bar{D}} C_{2} & D_{12} \left( I + \textcolor{MatlabGreen}{D_{c}} \textcolor{MatlabGreen}{\bar{D}} D_{22} \right) \textcolor{MatlabGreen}{C_{c}}
\end{bmatrix}, \\
\textcolor{MatlabGreen}{\mathcal{D}} &= D_{11} + D_{12} \textcolor{MatlabGreen}{D_{c}} \textcolor{MatlabGreen}{\bar{D}} D_{21},
\end{align}
\end{subequations}
%\begin{equation}
%\left[\begin{array}{c|c}
%\textcolor{MatlabGreen}{\mathcal{A}} & \textcolor{MatlabGreen}{\mathcal{B}} \\ \hline \textcolor{MatlabGreen}{\mathcal{C}} & \textcolor{MatlabGreen}{\mathcal{D}}
%\end{array} \right] = \left[\begin{array}{cc|c}
%A + B_{2} \textcolor{MatlabGreen}{D_{c}} \textcolor{MatlabGreen}{\bar{D}} C_{2} & B_{2} \left( I + \textcolor{MatlabGreen}{D_{c}} \textcolor{MatlabGreen}{\bar{D}} D_{22} \right) \textcolor{MatlabGreen}{C_{c}} & B_{1} + B_{2} \textcolor{MatlabGreen}{D_{c}} \textcolor{MatlabGreen}{\bar{D}} D_{21} \\ \textcolor{MatlabGreen}{B_{c}} \textcolor{MatlabGreen}{\bar{D}} C_{2} & \textcolor{MatlabGreen}{A_{c}} + \textcolor{MatlabGreen}{B_{c}} \textcolor{MatlabGreen}{\bar{D}} D_{22} \textcolor{MatlabGreen}{C_{c}} & \textcolor{MatlabGreen}{B_{c}} \textcolor{MatlabGreen}{\bar{D}} D_{21} \\ \hline
%C_{1} + D_{12} \textcolor{MatlabGreen}{D_{c}} \textcolor{MatlabGreen}{\bar{D}} C_{2} & D_{12} \left( I + \textcolor{MatlabGreen}{D_{c}} \textcolor{MatlabGreen}{\bar{D}} D_{22} \right) \textcolor{MatlabGreen}{C_{c}} & D_{11} + D_{12} \textcolor{MatlabGreen}{D_{c}} \textcolor{MatlabGreen}{\bar{D}} D_{21}
%\end{array} \right],
%\end{equation}
where $\textcolor{MatlabGreen}{\bar{D}} = \left( I - D_{22} \textcolor{MatlabGreen}{D_{c}} \right)^{-1}$. 
\subsection{Change of Variables}
The system is nonlinear in $\textcolor{MatlabGreen}{A_{c}}$, $\textcolor{MatlabGreen}{B_{c}}$, $\textcolor{MatlabGreen}{C_{c}}$, and $\textcolor{MatlabGreen}{D_{c}}$. To obtain an affine relation, the following change of variables is deployed
\begin{subequations}
\begin{align} \label{eq:CoV}
\textcolor{MatlabGreen}{A_{c2}} &= \textcolor{MatlabGreen}{A_{c}} + \textcolor{MatlabGreen}{B_{c}} \textcolor{MatlabGreen}{\bar{D}} D_{22} \textcolor{MatlabGreen}{C_{c}}, \\
\textcolor{MatlabGreen}{B_{c2}} &= \textcolor{MatlabGreen}{B_{c}} \textcolor{MatlabGreen}{\bar{D}}, \\
\textcolor{MatlabGreen}{C_{c2}} &= \left( I + \textcolor{MatlabGreen}{D_{c}} \textcolor{MatlabGreen}{\bar{D}} D_{22} \right) \textcolor{MatlabGreen}{C_{c}}, \\
\textcolor{MatlabGreen}{D_{c2}} &= \textcolor{MatlabGreen}{D_{c}} \textcolor{MatlabGreen}{\bar{D}},
\end{align}
\end{subequations}
%I.e.,
%\begin{equation}
%\left[\begin{array}{c|c}
%\textcolor{MatlabGreen}{A_{c2}} & \textcolor{MatlabGreen}{B_{c2}} \\ \hline \textcolor{MatlabGreen}{C_{c2}} & \textcolor{MatlabGreen}{D_{c2}}
%\end{array} \right] = \left[\begin{array}{c|c}
%\textcolor{MatlabGreen}{A_{c}} + \textcolor{MatlabGreen}{B_{c}} \left( I - D_{22} \textcolor{MatlabGreen}{D_{c}} \right)^{-1} D_{22} \textcolor{MatlabGreen}{C_{c}} & \textcolor{MatlabGreen}{B_{c}} \left( I - D_{22} \textcolor{MatlabGreen}{D_{c}} \right)^{-1} \\ \hline
%\left( I - \textcolor{MatlabGreen}{D_{c}} D_{22} \right)^{-1} \textcolor{MatlabGreen}{C_{c}} & \textcolor{MatlabGreen}{D_{c}} \left( I - D_{22} \textcolor{MatlabGreen}{D_{c}} \right)^{-1}
%\end{array} \right],
%\end{equation}
which renders the closed-loop system matrices
\begin{subequations}
\begin{align} \label{eq:CoV_result}
\textcolor{MatlabGreen}{\mathcal{A}} &= \begin{bmatrix}
A + B_{2} \textcolor{MatlabGreen}{D_{c2}} C_{2} & B_{2} \textcolor{MatlabGreen}{C_{c2}} \\
\textcolor{MatlabGreen}{B_{c2}} C_{2} & \textcolor{MatlabGreen}{A_{c2}}
\end{bmatrix}, \\
\textcolor{MatlabGreen}{\mathcal{B}} &= \begin{bmatrix} B_{1} + B_{2} \textcolor{MatlabGreen}{D_{c2}} D_{21} \\ \textcolor{MatlabGreen}{B_{c2}} D_{21} 
\end{bmatrix}, \\
\textcolor{MatlabGreen}{\mathcal{C}} &= \begin{bmatrix}
C_{1} + D_{12} \textcolor{MatlabGreen}{D_{c2}} C_{2} & D_{12} \textcolor{MatlabGreen}{C_{c2}}
\end{bmatrix}, \\ 
\textcolor{MatlabGreen}{\mathcal{D}} &= D_{11} + D_{11} + D_{12} \textcolor{MatlabGreen}{D_{c2}} D_{21}, 
\end{align}
\end{subequations}
affine in $\textcolor{MatlabGreen}{A_{c2}}$, $\textcolor{MatlabGreen}{B_{c2}}$, $\textcolor{MatlabGreen}{C_{c2}}$, and $\textcolor{MatlabGreen}{D_{c2}}$.
%\begin{rem}
%Note that the closed-loop system can equivalently be written as
%\begin{equation}
%\begin{split}
%& \left[\begin{array}{c|c}
%\textcolor{MatlabGreen}{\mathcal{A}} & \textcolor{MatlabGreen}{\mathcal{B}} \\ \hline \textcolor{MatlabGreen}{\mathcal{C}} & \textcolor{MatlabGreen}{\mathcal{D}}
%\end{array} \right] = \left[\begin{array}{cc|c}
%A & 0 & B_{1} \\ 0 & 0 & 0 \\ \hline
%C_{1} & 0 & D_{11} 
%\end{array} \right]\\ &\:\:+ \left[\begin{array}{cc}
%0 & B_{2} \\ I & 0 \\ \hline
%0 & D_{12} 
%\end{array} \right]
%\left[\begin{array}{cc}
%\textcolor{MatlabGreen}{A_{c2}} & \textcolor{MatlabGreen}{B_{c2}} \\ \textcolor{MatlabGreen}{C_{c2}} & \textcolor{MatlabGreen}{D_{c2}} 
%\end{array}\right]
%\left[\begin{array}{cc|c}
%0 & I & 0 \\ C_{2} & 0 & D_{21}
%\end{array} \right], \label{eq:22ish}
%\end{split}
%\end{equation}
%and that in case $D_{22} = 0$, it holds that $\textcolor{MatlabGreen}{A_{c}} = \textcolor{MatlabGreen}{A_{c2}}$, $\textcolor{MatlabGreen}{B_{c}} = \textcolor{MatlabGreen}{B_{c2}}$, $\textcolor{MatlabGreen}{C_{c}} = \textcolor{MatlabGreen}{C_{c2}}$, and $\textcolor{MatlabGreen}{D_{c}} = \textcolor{MatlabGreen}{D_{c2}}$.
%\end{rem}
The reverse change of variables is
\begin{subequations}
\begin{align}
\textcolor{MatlabGreen}{A_{c}} &= \textcolor{MatlabGreen}{A_{c2}} - \textcolor{MatlabGreen}{B_{c2}} D_{22} \left( I + \textcolor{MatlabGreen}{D_{c2}} D_{22} \right)^{-1} \textcolor{MatlabGreen}{C_{c2}}, \\
\textcolor{MatlabGreen}{B_{c}} &= \textcolor{MatlabGreen}{B_{c2}} \left( I + D_{22} \textcolor{MatlabGreen}{D_{c2}} \right)^{-1}, \\
\textcolor{MatlabGreen}{C_{c}} &= \left( I + \textcolor{MatlabGreen}{D_{c2}} D_{22} \right)^{-1} \textcolor{MatlabGreen}{C_{c2}}, \\
\textcolor{MatlabGreen}{D_{c}} &= \left( I + \textcolor{MatlabGreen}{D_{c2}} D_{22} \right)^{-1} \textcolor{MatlabGreen}{D_{c2}}.
\end{align} \label{eq:23ding}
\end{subequations}
To reconstruct the filter, the change of variables must be invertible, i.e., $\left( I + \textcolor{MatlabGreen}{D_{c2}} D_{22} \right)$ must be nonsingular. 

\subsection{Selecting Channels and Imposing Objectives} \label{subsec:Objective}
The objective is to compute a dynamical filter that meets various specifications on the closed-loop system behavior $T: \tilde{w} \rightarrow \tilde{z}$. Typically, these specifications are defined for particular channels $T_{j}: \tilde{w}_{j} \rightarrow \tilde{z}_{j}$ or combinations of channels. Specification $j$ of the objective is formulated relative to the closed-loop transfer function of the form
\begin{equation}
T_{j} = L_{j} T R_{j},
\end{equation}
where the matrices $L_{j}$, $R_{j}$ select the appropriate input/output (I/O) channels or channel combinations. I.e., this merely boils down to $w = R_{j} w_{j}$ and $z_{j} = L_{j} z$. Hence, $T_{j}$ admits the realization 
\begin{subequations}
\begin{align}
\textcolor{MatlabGreen}{\mathcal{A}} &= \begin{bmatrix}
A + B_{2} \textcolor{MatlabGreen}{D_{c2}} C_{2} & B_{2} \textcolor{MatlabGreen}{C_{c2}} \\
\textcolor{MatlabGreen}{B_{c2}} C_{2} & \textcolor{MatlabGreen}{A_{c2}}
\end{bmatrix}, \\ 
\textcolor{MatlabGreen}{\mathcal{B}_{j}} &= \begin{bmatrix}
B_{1,j} + B_{2} \textcolor{MatlabGreen}{D_{c2}} D_{21,j} \\ \textcolor{MatlabGreen}{B_{c2}} D_{21,j} 
\end{bmatrix}, \\
\textcolor{MatlabGreen}{\mathcal{C}_{j}} &= \begin{bmatrix}
C_{1,j} + D_{12,j} \textcolor{MatlabGreen}{D_{c2}} C_{2} & D_{12,j} \textcolor{MatlabGreen}{C_{c2}}
\end{bmatrix}, \\
\textcolor{MatlabGreen}{\mathcal{D}_{j}} &= D_{11,j} + D_{12,j} \textcolor{MatlabGreen}{D_{c2}}  D_{21,j},  
\end{align} \label{eq:LMI_Realization}
\end{subequations}
where $B_{1,j} := B_{1} R_{j}$,  $C_{1,j} := L_{j} C_{1}$,  $D_{11,j} := L_{j} D_{11} R_{j}$, $D_{12,j} := L_{j} D_{12}$, and $D_{21,j} := D_{21} R_{j}$.

The LMI approach expresses each specification or objective as a constraint on the closed-loop transfer functions $T_{j}(s)$ with a realization described by \Cref{eq:LMI_Realization}. Various objectives can be imposed on the isolated channels, see \cite{schererMultiobjectiveOutputfeedbackControl1997} for details. In particular, the minimum and a maximum gain constraint are imposed on the to be selected channels for fault diagnosis.

%\newpage

%===============================================================================
\section{Design Specifications and Synthesis for Fault Diagnosis} \label{sec:DESIGN SPECIFICATIONS AND SYNTHESIS FOR FAULT DIAGNOSIS}
Next, the fault diagnosis problem is considered. To this end, the setting depicted in \Cref{fig:CL_System} is considered, with a generalized disturbance channel consisting of disturbances $\tilde{d}$ and faults $\tilde{f}$, i.e., $\tilde{w} = \begin{bmatrix} \tilde{d}^{\top} & \tilde{f}^{\top} \end{bmatrix}^{\top}$. The generalized performance channel consists of the residual, i.e., $\tilde{z} = \varepsilon$. First, the $\mathcal{H}_{-} / \mathcal{H}_{\infty}$ specifications are defined. Subsequently, weighting filters are introduced for direct shaping of the singular values and it is illustrated how to deal with strictly proper systems. Next, the problem is posed as an optimization problem and the solution is presented in terms of matrix inequalities. Finally, the synthesis procedure is outlined.

\begin{defi}
Consider the system \eqref{eq:Ptilde_eq} and $\gamma >0$, $\nu > 0$. The fault diagnosis filter $Q$ is said to satisfy $\mathcal{H}_{-} / \mathcal{H}_{\infty}$ specifications if
\begin{enumerate}
\item $Q$ is proper and asymptotically stable; \label{cond1}
\item $\norm{T_{\epsilon \tilde{f}} (s)}_{-} > \nu$; \label{cond2}
\item $\norm{T_{\epsilon \tilde{d}} (s)}_{\infty} < \gamma$. \label{cond3}
\end{enumerate}
The objective considered in this paper is to find an admissible residual generator $Q$ which minimizes the sensitivity to disturbance $\gamma$, while simultaneously maximizing the sensitivity to faults $\nu$.
\end{defi}

Various mixed $\mathcal{H}_{-} / \mathcal{H}_{\infty}$ performance criteria can be considered, see, e.g., \cite{dingUnifiedApproachOptimization2000,wangLMIApproachIndex2007,henryTheoriesDesignAnalysis2021}. It is clear that better performance is achieved when the gap $\mathcal{J} := \frac{\nu}{\gamma}$ increases. In this paper, the ratio $\mathcal{J}$ is indirectly maximized through maximizing $\nu$, while constraining $\gamma$. Hence, the formal optimization problem is defined as 
\begin{equation}
\begin{split}
\max_{Q \in \mathcal{RH}_{\infty}, X \succ 0, \nu > 0, \gamma = \gamma_{0}} & \nu, \\
\mathrm{subject \: to} \qquad  & \eqref{eq:BMI_maxgain}, \eqref{eq:BMI_mingain},
\end{split} \label{eq:Optimprob}
\end{equation}
where the maximum disturbance gain $\gamma = \gamma_{0}$ is set.

\begin{rem}
Typically, the residual generator is connected to the system in open loop, $\tilde{z} = \tilde{u}$, which implies that the residual is directly fed through $P$. For that reason, scaling the residual generator, scales $\norm{T_{\epsilon \tilde{f}} (s)}_{-}$ and $\norm{T_{\epsilon \tilde{d}} (s)}_{\infty}$ equally, leaving the ratio $\mathcal{J}$, and thus performance, unchanged.
\end{rem}
\begin{rem}
A common Lyapunov variable $X$ in both the Bounded Real Lemma and the Minimum Gain Lemma enables to restrict the order of the resulting filter $Q$. By alleviating this constraint, additional freedom may result in a lower criterion.
\end{rem}
\begin{rem}
Note that stability of the overall system is embedded in the Bounded Real Lemma \eqref{eq:BMI_maxgain}.
\end{rem}
The Bounded Real Lemma and Minimum Gain Lemma are both a function of the closed-loop matrices and are transformed into the inequalities for synthesis next.

\subsection{Direct Shaping of Singular Values}
Next to bounding the singular values of particular channels, the singular values can also be shaped. Consider for instance invertible shaping filters on the generalized disturbance channel $G_{w_{j}}: 
\tilde{w}_{j} \rightarrow w_{j}$. In particular with diagonal shaping filters $G_{f}: \tilde{f} \rightarrow f$ and $G_{d}: \tilde{d} \rightarrow d$, specification (\ref{cond2}) and (\ref{cond3}) can be written as $\norm{T_{\epsilon f} (s) G_{f} (s)}_{-} > \nu$ and $\norm{T_{\epsilon d} (s) G_{d} (s)}_{\infty} < \gamma$ and equivalently as $\abs{T_{\epsilon f_{j}} (j \omega)} > \tfrac{\nu}{\abs{G_{f_{j}} (j \omega)}}$ $\forall \omega$ and $\abs{T_{\epsilon d_{j}} (j \omega)} < \tfrac{\gamma}{\abs{G_{d_{j}} (j \omega)}} \:\: \forall\omega$. From the latter property follows that the inverse of $G_{f}$ can be used to shape the minimum singular value of $T_{\varepsilon f}$ and the inverse of $G_{d}$ can be used to shape the maximum singular value of $T_{\varepsilon d}$.

\subsection{Strictly Proper Systems}
The minimum gain used in specification (\ref{cond2}) is zero for strictly proper systems, see Lemma \ref{lem:mingain}. This results in infeasibility of \Cref{eq:Optimprob}. With an accommodation using shaping filters, the proposed method can still be applied at the cost of reduced fault sensitivity at higher frequencies.

Suppose that the output of the filter $Q$ directly forms the residual, i.e., $\tilde{z} = \tilde{u}$. Since $Q$ should be implementable, i.e., proper and stable, the transfer $T_{\tilde{y} \tilde{w_{j}}}$ must be proper and stable. If a weighting filter is present in the latter, e.g., $T_{\tilde{y} \tilde{w_{j}}} = T_{\tilde{y} w_{j}} G_{w_{j}}$, the transfer function can be made proper with appropriate choice of improper $G_{w_{j}}$. Since in that case $G_{w_{j}}^{-1}$ goes to zero for high frequency, fault sensitivity at higher frequencies is lost as the minimum gain constraint is relaxed at these frequencies.

\subsection{Synthesis Matrix Inequalities}
First, the transformed result is stated, after which the transformed optimization problem is posed.
%The following two statements are equivalent
\begin{thm}
If there exist $\nu > 0$, $\gamma > 0$, $X_{1} = X_{1}^{\top} \succ 0$, $Y_{1} = Y_{1}^{\top} \succ 0$, $A_{n}$, $B_{n}$, $C_{n}$, $D_{n}$, $\mathcal{X}$, $\mathcal{Y}$, $\mathcal{Z}$ such that the maximum gain LMI
\begin{equation}
\begin{bmatrix} M_{11} & M_{12} & M_{13} & M_{14} \\ * & M_{22} & M_{23} & M_{24} \\ * & * & M_{33} & M_{34} \\ * & * & * & M_{44} \end{bmatrix} \prec 0, \label{eq:EQ1_Syn}
\end{equation}
the minimum gain bilinear matrix inequality (BMI)
\begin{equation}
\begin{bmatrix} N_{11} & N_{12} & N_{13} & N_{14} \\ * & N_{22} & N_{23} & N_{24} \\ * & * & N_{33} & N_{34} \\ * & * & * & N_{44} \end{bmatrix} \prec 0, \label{eq:EQ2_Syn}
\end{equation}
and
\begin{equation}
\begin{bmatrix}
\textcolor{MatlabGreen}{X_{1}} & I \\
* & \textcolor{MatlabGreen}{Y_{1}} \end{bmatrix}
 \succ 0, \label{eq:EQ3_Syn}
\end{equation}
where the entries of $M$ and $N$ are given in Appendix \ref{sec:APPENDIX: SYNTHESIS INEQUALITIES}, hold, then $Q = \left[\begin{array}{c|c}
\textcolor{MatlabGreen}{\mathcal{A}_{c}} & \textcolor{MatlabGreen}{\mathcal{B}_{c}} \\ \hline \textcolor{MatlabGreen}{\mathcal{C}_{c}} & \textcolor{MatlabGreen}{\mathcal{D}_{c}}
\end{array} \right] \in \mathcal{RH}_{\infty}$ exists such that $\norm{T_{\epsilon \tilde{f}} (s)}_{-} > \nu$ and $\norm{T_{\epsilon \tilde{d}} (s)}_{\infty} < \gamma$.
\end{thm}

A brief outline of the proof is given below. The full proof will be published elsewhere due to space limitations.

\begin{proof-wo}
According to the transformation lemma, there exists a matrix completion $X_{2}$, $Y_{2}$, $X_{3}$, $Y_{3}$ and a half dual variable $Y_{\mathrm{CL}}$ which is full rank. With the matrix completion and by definition,
\begin{equation}
\begin{split}
&\left[\begin{array}{cc}
\textcolor{MatlabGreen}{A_{c2}} & \textcolor{MatlabGreen}{B_{c2}} \\ \textcolor{MatlabGreen}{C_{c2}} & \textcolor{MatlabGreen}{D_{c2}} 
\end{array}\right] = \begin{bmatrix}
\textcolor{MatlabGreen}{X_{2}} & \textcolor{MatlabGreen}{X_{1}} B_{2} \\ 0 & I
\end{bmatrix}^{-1} \\ &\qquad  \left( \begin{bmatrix}
\textcolor{MatlabGreen}{A_{n}} & \textcolor{MatlabGreen}{B_{n}} \\ \textcolor{MatlabGreen}{C_{n}} & \textcolor{MatlabGreen}{D_{n}} 
\end{bmatrix} - \begin{bmatrix}
\textcolor{MatlabGreen}{X_{1}} A \textcolor{MatlabGreen}{Y_{1}} & 0 \\ 0 & 0
\end{bmatrix} \right)  \begin{bmatrix}
\textcolor{MatlabGreen}{Y_{2}}^{\top} & 0 \\ C_{2} \textcolor{MatlabGreen}{Y_{1}} & I
\end{bmatrix}^{-1}.
\end{split} \label{eq:lastres}
\end{equation}
From this, the filter $Q$ can be reconstructed through the reverse change of variables \eqref{eq:23ding}. Writing \cref{eq:LMI_Realization} in the form of \cref{eq:23ding} and substitution of \cref{eq:lastres} gives a description of the particular isolated closed-loop channel.

Considering the maximum gain constraint first, the aim is to show that the resulting description satisfies $\norm{T_{1} (s)}_{\infty} < \gamma$, where $T_{1}(s)$ denotes the transfer function corresponding to the realization $\left( \textcolor{MatlabGreen}{\mathcal{A}}, \textcolor{MatlabGreen}{\mathcal{B}_{1}}, \textcolor{MatlabGreen}{\mathcal{C}_{1}}, \textcolor{MatlabGreen}{\mathcal{D}_{1}} \right)$. Substitution of this realization into the Bounded Real Lemma and applying a congruence transformation with $\mathrm{diag}\left( Y_{\mathrm{CL}}^{\top}, I, I \right)$ gives
\begin{equation}
\resizebox{\columnwidth}{!}{%
$ \begin{bmatrix}
\textcolor{MatlabGreen}{Y_{\mathrm{CL}}}^{\top} \textcolor{MatlabGreen}{\mathcal{A}}^{\top} \textcolor{MatlabGreen}{X_{\mathrm{CL}}}^{\top} + \textcolor{MatlabGreen}{X_{\mathrm{CL}}} \textcolor{MatlabGreen}{\mathcal{A}} \textcolor{MatlabGreen}{Y_{\mathrm{CL}}} & \textcolor{MatlabGreen}{X_{\mathrm{CL}}} \textcolor{MatlabGreen}{\mathcal{B}_{1}} & \textcolor{MatlabGreen}{Y_{\mathrm{CL}}}^{\top} \textcolor{MatlabGreen}{\mathcal{C}_{1}}^{\top} \\
* & -\textcolor{MatlabRed}{\gamma^{2}} I & \textcolor{MatlabGreen}{\mathcal{D}_{1}}^{\top } \\
* & * & - I
\end{bmatrix} \prec 0,$} \label{eq:BRLsubs}
\end{equation}
where $X_{\mathrm{CL}} = X Y_{\mathrm{CL}}$. After substituting all the variables, it can be shown that \eqref{eq:BRLsubs} is equivalent to \eqref{eq:EQ1_Syn} and thus proving that indeed $\norm{T_{1} (s)}_{\infty} > \gamma$.

Next, consider the minimum gain constraint. Similarly, the aim is to show that the resulting description satisfies $\norm{T_{2} (s)}_{-} > \nu$, where $T_{2}(s)$ denotes the transfer function corresponding to the realization $\left( \textcolor{MatlabGreen}{\mathcal{A}}, \textcolor{MatlabGreen}{\mathcal{B}_{2}}, \textcolor{MatlabGreen}{\mathcal{C}_{2}}, \textcolor{MatlabGreen}{\mathcal{D}_{2}} \right)$. Substitution into the Minimum Gain Lemma after taking the Schur complement gives
\begin{equation}
\begin{bmatrix}
\textcolor{MatlabGreen}{X} \textcolor{MatlabGreen}{\mathcal{A}} + \textcolor{MatlabGreen}{\mathcal{A}}^{\top} \textcolor{MatlabGreen}{X} - \textcolor{MatlabGreen}{\mathcal{C}}_{2}^{\top} \textcolor{MatlabGreen}{\mathcal{C}}_{2} & \textcolor{MatlabGreen}{X} \textcolor{MatlabGreen}{\mathcal{B}}_{2} - \textcolor{MatlabGreen}{\mathcal{C}}^{\top}_{2} \textcolor{MatlabGreen}{\mathcal{D}}_{2}  \\
\textcolor{MatlabGreen}{\mathcal{B}}^{\top}_{2} \textcolor{MatlabGreen}{X}  - \textcolor{MatlabGreen}{\mathcal{D}}^{\top}_{2} \textcolor{MatlabGreen}{\mathcal{C}}_{2}  & \textcolor{MatlabRed}{\nu}^{2} I - \textcolor{MatlabGreen}{\mathcal{D}}^{\top}_{2} \textcolor{MatlabGreen}{\mathcal{D}} _{2}
\end{bmatrix} < 0,
\end{equation}
which can be written as
\begin{equation}
\begin{bmatrix}
\textcolor{MatlabGreen}{X} \textcolor{MatlabGreen}{\mathcal{A}} + \textcolor{MatlabGreen}{\mathcal{A}}^{\top} \textcolor{MatlabGreen}{X} & \textcolor{MatlabGreen}{X} \textcolor{MatlabGreen}{\mathcal{B}} \\
\textcolor{MatlabGreen}{\mathcal{B}}^{\top} \textcolor{MatlabGreen}{X} & \textcolor{MatlabRed}{\nu}^{2} I 
\end{bmatrix} < \begin{bmatrix}
\textcolor{MatlabGreen}{\mathcal{C}}^{\top} \\
\textcolor{MatlabGreen}{\mathcal{D}}^{\top}
\end{bmatrix}   \begin{bmatrix}
\textcolor{MatlabGreen}{\mathcal{C}} & \textcolor{MatlabGreen}{\mathcal{D}} 
\end{bmatrix}.
\end{equation}
Using Youngs relation with $F = \begin{bmatrix}
\textcolor{MatlabGreen}{\mathcal{C}} & \textcolor{MatlabGreen}{\mathcal{D}} 
\end{bmatrix}$ and auxiliary slack variables $G = \begin{bmatrix}
\textcolor{MatlabBlue}{\mathcal{Y}} & \textcolor{MatlabBlue}{\mathcal{X}}
\end{bmatrix}$, and thereafter the Schur complement,
\begin{equation}
\resizebox{\columnwidth}{!}{%
$\begin{bmatrix}
\textcolor{MatlabGreen}{X} \textcolor{MatlabGreen}{\mathcal{A}} + \textcolor{MatlabGreen}{\mathcal{A}}^{\top} \textcolor{MatlabGreen}{X} - \textcolor{MatlabGreen}{\mathcal{C}}^{\top} \textcolor{MatlabBlue}{\mathcal{Y}} - \textcolor{MatlabBlue}{\mathcal{Y}}^{\top} \textcolor{MatlabGreen}{\mathcal{C}}  & \textcolor{MatlabGreen}{X} \textcolor{MatlabGreen}{\mathcal{B}} - \textcolor{MatlabGreen}{\mathcal{C}}^{\top} \textcolor{MatlabBlue}{\mathcal{X}} - \textcolor{MatlabBlue}{\mathcal{Y}}^{\top} \textcolor{MatlabGreen}{\mathcal{D}} & \textcolor{MatlabBlue}{\mathcal{Y}}^{\top} \\
* & \textcolor{MatlabRed}{\nu^{2}} I - \textcolor{MatlabBlue}{\mathcal{X}}^{\top} \textcolor{MatlabGreen}{\mathcal{D}} - \textcolor{MatlabGreen}{\mathcal{D}}^{\top} \textcolor{MatlabBlue}{\mathcal{X}} & \textcolor{MatlabBlue}{\mathcal{X}}^{\top} \\
* & * & -I
\end{bmatrix} < 0,$} \label{eq:precongr}
\end{equation}
is obtained. Now applying the same congruence transformation with $\mathrm{diag}\left( Y_{\mathrm{CL}}^{\top}, I, I \right)$ and substituting all the variables, it can be shown that \eqref{eq:precongr} is equivalent to \eqref{eq:EQ2_Syn} and thus proving that indeed $\norm{T_{2} (s)}_{-} > \nu$. \QEDA
\end{proof-wo}
See also \cite{schererMultiobjectiveOutputfeedbackControl1997,caverlyHinftyOptimalParallel2018} for related results.

\begin{rem}
The minimum gain BMI can be used in conjunction with classical multiobjective LMI formulations such as generalized $\mathcal{H}_{2}$ performance, peak amplitude constraints, or regional pole constraints, since the same change of variables is employed, see \cite{schererMultiobjectiveOutputfeedbackControl1997} for details.
\end{rem}

The transformed optimization problem is then given by
\begin{equation}
\begin{split}
\max_{X_{1} \succ 0, Y_{1} \succ 0, \nu > 0, \gamma = \gamma_{0}, A_{n}, B_{n}, C_{n}, D_{n}, \mathcal{X}, \mathcal{Y}, \mathcal{Z}} & \nu, \\
\mathrm{subject \: to} \qquad \qquad \qquad & \Cref{eq:EQ1_Syn} 
\mathrm{\: to \:} \Cref{eq:EQ3_Syn}.
\end{split} \label{eq:optimprob}
\end{equation}

The minimum gain matrix inequality remains bilinear whereas the maximum gain inequality is affine in the free variables. For that reason the synthesis is performed iteratively through the following procedure.

\subsection{Synthesis} \label{subsec:Synthesis}
%\atodo{In algorithm format of niet?}
There are many methods to solve BMIs, see, e.g., \cite{hassibiPathfollowingMethodSolving1999}. The BMI is solved iteratively by solving sequential LMI problems, where each iterate is denoted by $k$. First, the variable $\gamma = \gamma_{0}$ is set.

\begin{enumerate}\addtocounter{enumi}{-1}
\item For $k=0$, set $\textcolor{MatlabBlue}{\mathcal{X}}_{k} = I$, $\textcolor{MatlabBlue}{\mathcal{Y}}_{k} = 0$, $\textcolor{MatlabBlue}{\mathcal{Z}}_{k} = 0$
\item Fix $\textcolor{MatlabBlue}{\mathcal{X}}_{k}$, $\textcolor{MatlabBlue}{\mathcal{Y}}_{k}$, $\textcolor{MatlabBlue}{\mathcal{Z}}_{k}$, and solve for $\textcolor{MatlabGreen}{A}_{n,k}$, $\textcolor{MatlabGreen}{B}_{n,k}$, $\textcolor{MatlabGreen}{C}_{n,k}$, $\textcolor{MatlabGreen}{D}_{n,k}$, $\textcolor{MatlabGreen}{X}_{1,k}=\textcolor{MatlabGreen}{X}_{1,k}^{\top} \succ 0$, $\textcolor{MatlabGreen}{Y}_{1,k}=\textcolor{MatlabGreen}{Y}_{1,k}^{\top} \succ 0$ and $0 < \nu_{k}^{2} < \infty$ such that $\mathcal{J} = \nu_{k}^{2}$ is maximized, subject to \eqref{eq:EQ1_Syn}, \eqref{eq:EQ2_Syn}, and \eqref{eq:EQ3_Syn}.
\item Fix the just obtained  $A_{n,k}$, $B_{n,k}$, $C_{n,k}$, $D_{n,k}$, $X_{1,k}=X_{1,k}^{\top} > 0$, $Y_{1,k}=Y_{1,k}^{\top} > 0$, and solve for $\textcolor{MatlabBlue}{\mathcal{X}}_{k}$, $\textcolor{MatlabBlue}{\mathcal{Y}}_{k}$, $\textcolor{MatlabBlue}{\mathcal{Z}}_{k}$ and $0 < \nu_{k}^{2} < \infty$ such that $\mathcal{J} = \nu_{k}^{2}$ is maximized, subject to \eqref{eq:EQ1_Syn}.
\item Return to 1 if $\abs{ \nu_{k-1}^{2} - \nu_{k}^{2} } > \mu$, where $\mu$ specifies a tolerance.
\end{enumerate}
Next, the proposed algorithm is applied to an example.

%%%%%%%%%%%%%%%%%%%%%%%%%%%%%%%%%%%%%%%%%%%%%%%%%%%%%%%%%%%%%%%%%%%%%%%%%%%%%%%%

\section{Numerical Example} \label{sec:NUMERICAL EXAMPLE}
The following example illustrates the proposed approach on a fault diagnosis problem. All LMI-related computations are performed with YALMIP \cite{lofbergYALMIPToolboxModeling2004} and solved with MOSEK \cite{apsMOSEKOptimizationToolbox}.
%
%\begin{figure}[t]
%\centering
%\includegraphics[]{Figs/CDC_Numerical_Example_Scheme_1}
%%		\tikzexternaldisable   % Comment when tikzexternalize is off
%\caption{Closed-loop controlled regulator problem, with plant $P$ and controller $C$, subjected to actuator faults $f$ and output disturbance $d$. The fault and disturbance are weighted with $G_{f}$, and $G_{d}$, respectively. The controlled system is augmented with the residual generator $Q$.}
%% 		\tikzexternalenable    % Comment when tikzexternalize is off
%\label{fig:CL_System1}
%\end{figure}
%
%
\begin{figure}[t]
\centering
\includegraphics[]{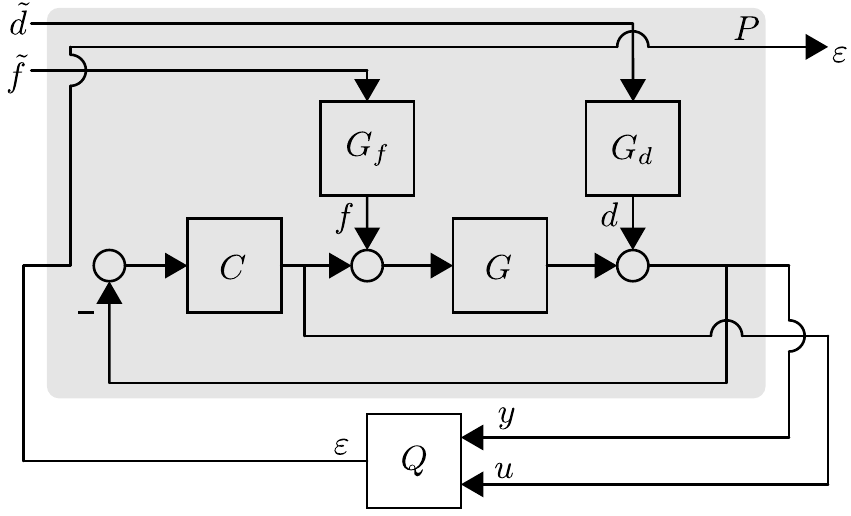}
%		\tikzexternaldisable   % Comment when tikzexternalize is off
\caption{Closed-loop controlled regulator problem in generalized plant form, c.f., \Cref{fig:CL_System}. The generalized plant is highlighted \tikzsquare{MatlabGray80}.}
% 		\tikzexternalenable    % Comment when tikzexternalize is off
\label{fig:CL_System3}
\end{figure}
% (s+25)(s+15)(s+5)/(s+10)/(s+40)/(s+3)
Consider the closed-loop control configuration in \Cref{fig:CL_System3}, with single-input single-output plant $G(s) = \frac{(s+25)(s+15)(s+5)}{(s+40)(s+10)(s+3)}$ and controller $C(s) = 15s + 25$. The system is subjected to disturbances $d$ and actuator faults $f$, which are weighted by $G_{d}(s)$ and $G_{f}(s)$, respectively. Consider $\tilde{w} := \begin{bmatrix} \tilde{d} & \tilde{f} \end{bmatrix}^{\top}$, $\tilde{z} := \varepsilon$, $\tilde{y} := \begin{bmatrix}y & u\end{bmatrix}^{\top}$, and $\tilde{u} := \varepsilon$. The transfer function matrix of the generalized plant from $\begin{bmatrix} \tilde{w} & \tilde{u} \end{bmatrix}^\top$ to $\begin{bmatrix}\tilde{z} & \tilde{y} \end{bmatrix}^\top$, see \Cref{fig:CL_System3}, is given by
\begin{equation}
\begin{bmatrix}
\tilde{z}\\ \hline
\tilde{y}
\end{bmatrix}=\left[
    \begin{array}{cc|l}
        0 & 0 & I \\ \hline
        S_{O} G_{d} & S_{O} G_{u} G_{f} & 0 \\
        -S_{I} C G_{d} & -S_{I} C G_{u} G_{f} & 0
    \end{array}
\right] \begin{bmatrix}
\tilde{w}\\ \hline
\tilde{u}
\end{bmatrix},
\end{equation}
where the input sensitivity $S_{I} = (I + C G )^{-1}$, and the output sensitivity $S_{O} = (I + G C )^{-1}$. The generalized disturbance input is split into $\tilde{w} = \begin{bmatrix} \tilde{w}_{\infty} & \tilde{w}_{-} \end{bmatrix}^{\top} = \begin{bmatrix} \tilde{d} & \tilde{f} \end{bmatrix}^{\top}$. It is aimed to impose $\mathcal{H}_{\infty}$ performance from $\tilde{w}_{\infty}$ to $\tilde{z}$ and impose $\mathcal{H}_{-}$ sensitivity from $\tilde{w}_{-}$ to $\tilde{z}$. Two arbitrary weighting filters are chosen to emphasize the shaping capability of the proposed method. To that end $G_{d} = \frac{s^{4} + 62.8 s^{3} + 1392 s^2 + 1.43 \cdot{10}^{4} s + 4.87 \cdot{10}^{4}}{s^{4} + 332 s^{3} + 2724 s^2 + 8.10 \cdot{10}^{4} s + 1.22 \cdot{10}^{5}}$ and $G_{f} = \frac{0.92 s^4 + 43.25 s^3 + 1911 s^2 + 5976 s + 1.75 \cdot{10}^{4}}{s^4 + 13.19 s^3 + 3966 s^2 + 2605 s + 3.90\cdot{10}^{4}}$.

%Consider the multiobjective optimization problem, described by \Cref{eq:optimprob}, which effectively maximizes
%\begin{equation}
%\begin{split}
%\max_{X_{1}, Y_{1} \succ 0, \nu > 0, \gamma = \gamma_{0}, A_{n}, B_{n}, C_{n}, D_{n}, \mathcal{X}, \mathcal{Y}, \mathcal{Z}} & \norm{T_{z w_{-}}}_{-} \\
%\mathrm{subject \: to} \qquad  \qquad \quad & \norm{T_{z w_{\infty}}} < \gamma_{0},
%\end{split}
%\end{equation}
%where $\gamma_{0}=10$ and the transfer matrices $T_{z w_{\infty}}$ and $T_{z w_{-}}$ are isolated via the procedure in \Cref{subsec:Objective}.

Consider the multiobjective optimization problem, described by \Cref{eq:optimprob}, where $\gamma_{0}=1$, $T = \mathcal{F}_{l}(P,Q) = \begin{bmatrix}
T_{\tilde{z} \tilde{w}_{\infty}} & T_{\tilde{z} \tilde{w}_{-}}
\end{bmatrix}^{\top} = \begin{bmatrix}
T_{\varepsilon \tilde{d}} & T_{\varepsilon \tilde{f}}
\end{bmatrix}^{\top}$, and the transfer matrices $T_{\tilde{z} \tilde{w}_{\infty}}$ and $T_{\tilde{z} \tilde{w}_{-}}$ admit the realization, described in \Cref{subsec:Objective}, and are used to impose the maximum and minimum gain constraint, respectively.

\begin{figure}[t]
\centering
\includegraphics[width=\linewidth]{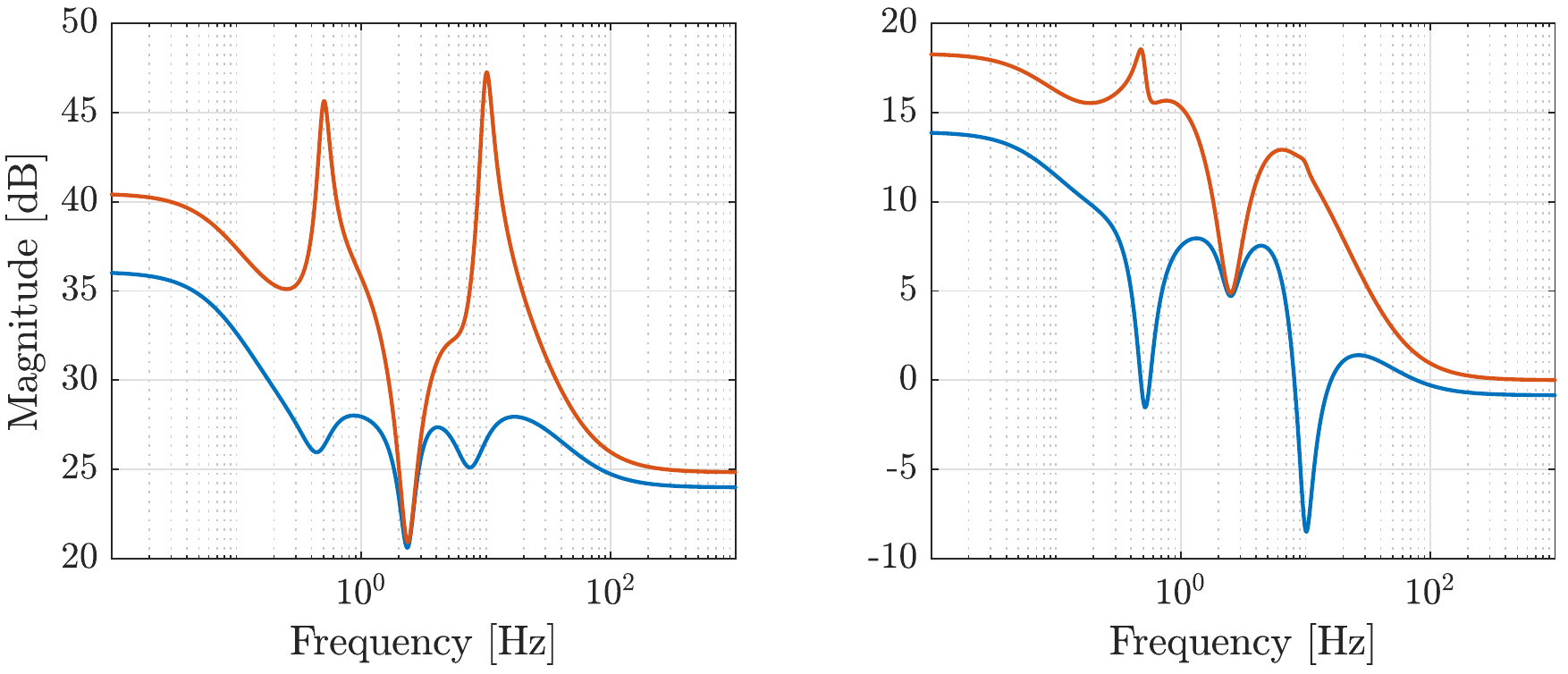}
%		\tikzexternaldisable   % Comment when tikzexternalize is off
\caption{Residual filter $Q : \begin{bmatrix}
y & u
\end{bmatrix}^{\top} \rightarrow \varepsilon$ with optimal ratio $\mathcal{J}$ \tikzline{MatlabBlue2}. The updated filter $Q_{2}$, with the same $\mathcal{J}$, is shown as \tikzline{MatlabRed2}.}
% 		\tikzexternalenable    % Comment when tikzexternalize is off
\label{fig:CL_System4}
\end{figure}

\begin{figure}[t]
\centering
\includegraphics[width=\linewidth]{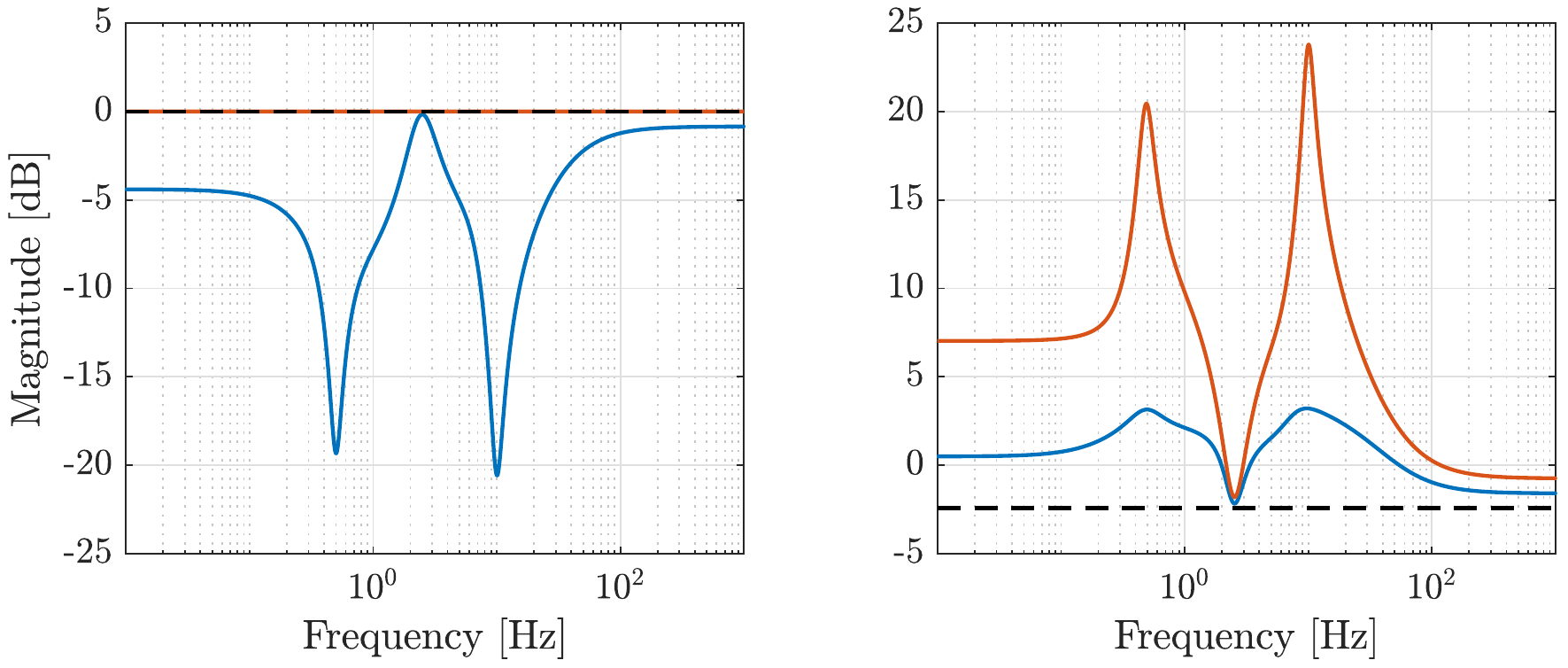}
%		\tikzexternaldisable   % Comment when tikzexternalize is off
\caption{Closed-loop system, i.e., $T : \begin{bmatrix}
\tilde{w}_{\infty} & \tilde{w}_{-}
\end{bmatrix}^{\top} \rightarrow \varepsilon$, depicted as \tikzline{MatlabBlue2}. The obtained bounds $\gamma$ and $\nu$ are indicated as \tikzdashedline{MatlabBlack}, showing tight bounds near $\omega = 2.5$ Hz. In this region, the fault to disturbance ratio is close to $\mathcal{J}$, whereas at other frequencies, fault sensitivity to disturbance attenuation ratio is higher. The closed-loop system with $Q_{2}$ is shown as \tikzline{MatlabRed2}.}
% 		\tikzexternalenable    % Comment when tikzexternalize is off
\label{fig:CL_System5}
\end{figure}

Solving this problem with the synthesis method, described in \Cref{subsec:Synthesis}, yields the optimum $\nu = 0.76$. So the ratio $\mathcal{J} = \frac{\nu}{\gamma} = 0.76$ and the obtained residual generator $Q$ is depicted in \Cref{fig:CL_System4}. Hence, with this particular residual generator it is guaranteed that there is disturbance suppression $\norm{T_{\varepsilon \tilde{d}}}_{\infty} < 1$ and fault sensitivity $\norm{T_{\varepsilon \tilde{f}}}_{-} > 0.76$, see \Cref{fig:CL_System5}. Additionally, the channels $T_{\varepsilon d}$ and $T_{\varepsilon f}$ are depicted with their found bounds $\gamma G_{d}^{-1}$ and  $\nu G_{f}^{-1}$.

\begin{figure}[t]
\centering
\includegraphics[width=\linewidth]{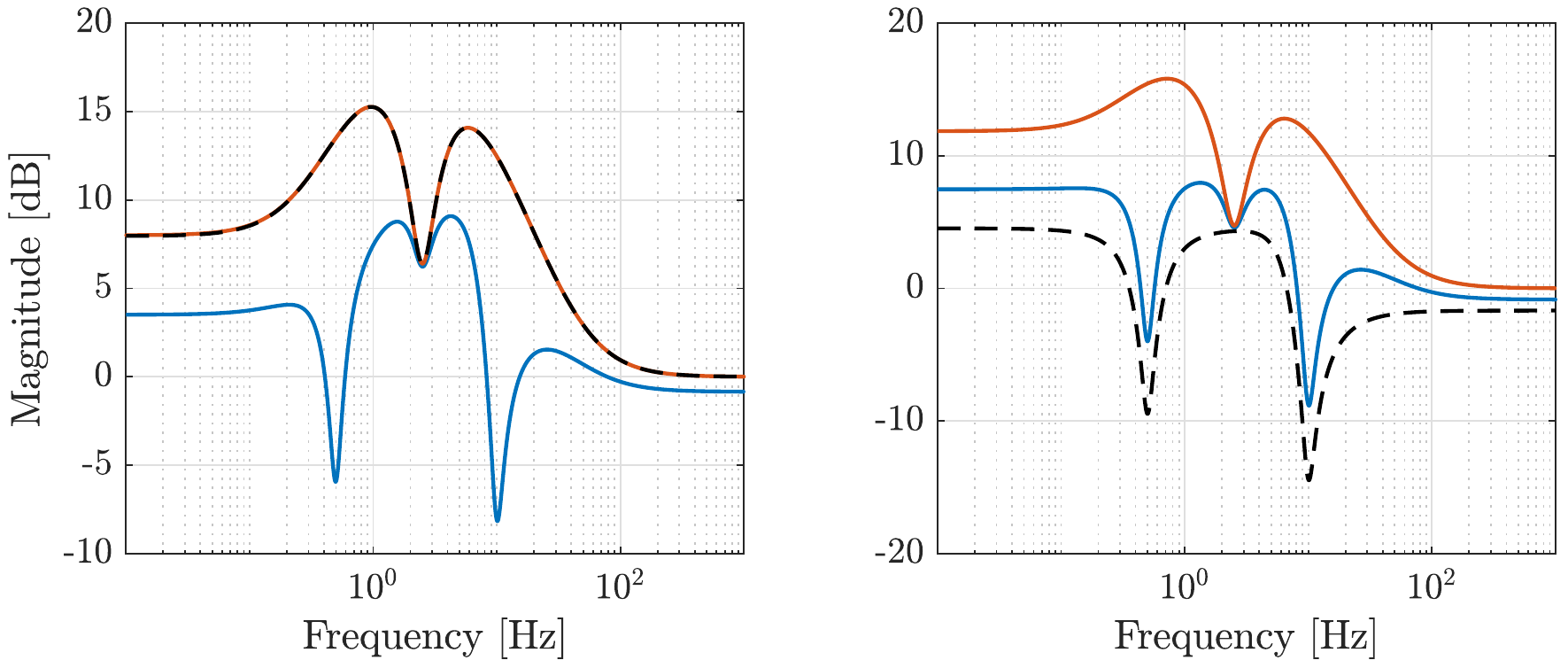}
%		\tikzexternaldisable   % Comment when tikzexternalize is off
\caption{Closed-loop system without weighting filters, i.e., $T_{\varepsilon d}$ and $T_{\varepsilon f}$, depicted as \tikzline{MatlabBlue2}. The obtained bounds $\gamma G_{d}^{-1}$ and $\nu G_{f}^{-1}$ are depicted as \tikzdashedline{MatlabBlack}. Indeed, a filter $Q$ is found that satisfies the shaped upperbound and lowerbound. The closed-loop system with $Q_{2}$ is shown as \tikzline{MatlabRed2}.}
% 		\tikzexternalenable    % Comment when tikzexternalize is off
\label{fig:CL_System6}
\end{figure}
From \Cref{fig:CL_System5} and \Cref{fig:CL_System6} can be concluded that there is still margin to improve the fault sensitivity while simultaneously satisfying $\norm{T_{\varepsilon \tilde{d}}}_{\infty} < 1$. Updating the filter $Q$ with this difference, i.e. $Q_{2} = \tfrac{\gamma G_{d}^{-1}}{ T_{\varepsilon d}} Q_{1}$, yields equivalent results to \cite{liuOptimalSolutionsMultiobjective2007}. Note that the objective $\mathcal{J} = 0.76$ remains equal.

%%%%%%%%%%%%%%%%%%%%%%%%%%%%%%%%%%%%%%%%%%%%%%%%%%%%%%%%%%%%%%%%%%%%%%%%%%%%%%%%

\section{Conclusion} \label{sec:CONCLUSION}
In this paper, a new method to solve the $\mathcal{H}_{-}/\mathcal{H}_{\infty}$ problem is presented. In particular, a method is proposed to shape the minimum and maximum singular value of the closed-loop performance channel and its effectiveness is illustrated in the context of fault diagnosis. A bilinear matrix inequality is derived which can directly be implemented in combination with various multiobjective matrix inequalities and applied to synthesize filters for a wide range of control and estimation problems.

%%%%%%%%%%%%%%%%%%%%%%%%%%%%%%%%%%%%%%%%%%%%%%%%%%%%%%%%%%%%%%%%%%%%%%%%%%%%%%%%
\section{Acknowledgements}
This work is supported by Topconsortia voor Kennis en Innovatie (TKI), and is supported by ASML Research, Veldhoven, The Netherlands.

%\textcolor{MatlabRed}{Check citaties}

\renewcommand*{\bibfont}{\footnotesize}
\bibliography{D:/SURFdrive/Zotero/KHJ_Classens_Zotero_Library_Better_BibTeX.bib}

%\bibliography{Zotero_Library/2022_SAFEPROCESS.bib}
%%\addbibresource{C:/Users/s137166/surfdrive/Documents/Zotero/KHJ_Classens_Zotero_Library.bib}
%\printbibliography
%\small
%\printbibliography
%\newpage
\section{APPENDIX: Synthesis Inequalities} \label{sec:APPENDIX: SYNTHESIS INEQUALITIES}

%\newpage
%\subsection{Synthesis Inequalities} \label{sec:APPENDIX: SYNTHESIS INEQUALITIES}
The entries of the maximum gain synthesis LMI are
\begin{subequations}
\begin{align}
M_{11} &= A \textcolor{MatlabGreen}{Y_{1}} + \textcolor{MatlabGreen}{Y_{1}} A^{\top} + B_{2} \textcolor{MatlabGreen}{C_{n}} + \textcolor{MatlabGreen}{C_{n}}^{\top} B_{2}^{\top}, \\
M_{12} &= A +  \textcolor{MatlabGreen}{A_{n}}^{\top} + B_{2} \textcolor{MatlabGreen}{D_{n}} C_{2}, \\
M_{13} &= B_{1,j} + B_{2} \textcolor{MatlabGreen}{D_{n}} D_{21,j}, \\
M_{14} &= \textcolor{MatlabGreen}{Y_{1}} C_{1,j}^{\top} + \textcolor{MatlabGreen}{C_{n}}^{\top} D_{12,j}^{\top}, \\
M_{22} &= \textcolor{MatlabGreen}{X_{1}} A + A^{\top} \textcolor{MatlabGreen}{X_{1}} + \textcolor{MatlabGreen}{B_{n}} C_{2} + C_{2}^{\top} \textcolor{MatlabGreen}{B_{n}}^{\top}, \\
M_{23} &= \textcolor{MatlabGreen}{X_{1}} B_{1,j} + \textcolor{MatlabGreen}{B_{n}} D_{21,j}, \\
M_{24} &= C_{1,j}^{\top} + C_{2}^{\top} \textcolor{MatlabGreen}{D_{n}}^{\top} D_{12,j}^{\top}, \\
M_{33} &= -\textcolor{MatlabRed}{\gamma^{2}} I, \\
M_{34} &= D_{11,j}^{\top} + D_{21,j}^{\top} \textcolor{MatlabGreen}{D_{n}}^{\top} D_{12,j}^{\top}, \\
M_{44} &= -I.
\end{align}
\end{subequations}
%\newpage

The entries of the minimum gain synthesis BMI are
\begin{subequations}
\begin{align}
\begin{split}
N_{11} &= A \textcolor{MatlabGreen}{Y_{1}} +  B_{2} \textcolor{MatlabGreen}{C_{n}} + \textcolor{MatlabGreen}{Y_{1}} A^{\top} + \textcolor{MatlabGreen}{C_{n}}^{\top} B_{2}^{\top} \\ & \quad- \left( \textcolor{MatlabGreen}{Y_{1}} C_{1,j}^{\top} + \textcolor{MatlabGreen}{C_{n}}^{\top} D_{12,j}^{\top} \right)  \textcolor{MatlabBlue}{\mathcal{Y}}   \\ & \quad -  \textcolor{MatlabBlue}{\mathcal{Y}}^{\top}  \left( C_{1,j} \textcolor{MatlabGreen}{Y_{1}} + D_{12,j} \textcolor{MatlabGreen}{C_{n}} \right), 
\end{split} \\
\begin{split}
N_{12} &= A + \textcolor{MatlabGreen}{A_{n}}^{\top} + B_{2} \textcolor{MatlabGreen}{D_{n}} C_{2} \\ & \quad - \left( \textcolor{MatlabGreen}{Y_{1}} C_{1,j}^{\top} + \textcolor{MatlabGreen}{C_{n}}^{\top} D_{12,j}^{\top} \right) \textcolor{MatlabBlue}{\mathcal{Z}} \\ & \quad  -  \textcolor{MatlabBlue}{\mathcal{Y}}^{\top} \left( C_{1,j} + D_{12,j} \textcolor{MatlabGreen}{D_{n}} C_{2} \right), \end{split} \\
\begin{split}
N_{13} &= B_{1,j} + B_{2} \textcolor{MatlabGreen}{D_{n}} D_{21,j} \\ & \quad - \textcolor{MatlabBlue}{\mathcal{Y}}^{\top} \left( D_{11,j} + D_{12,j} \textcolor{MatlabGreen}{D_{n}} D_{21,j} \right) \\ & \quad - \left( \textcolor{MatlabGreen}{Y_{1}} C_{1,j}^{\top} + \textcolor{MatlabGreen}{C_{n}}^{\top} D_{12,j}^{\top} \right) \textcolor{MatlabBlue}{\mathcal{X}}, \end{split} \\
N_{14} &= \textcolor{MatlabBlue}{\mathcal{Y}}^{\top}, \\
\begin{split}
N_{22} &= \textcolor{MatlabGreen}{X_{1}} A + \textcolor{MatlabGreen}{B_{n}} C_{2} + A^{\top} \textcolor{MatlabGreen}{X_{1}} + C_{2}^{\top} \textcolor{MatlabGreen}{B_{n}}^{\top}\\ & \quad  - \left( C_{1,j}^{\top} + C_{2}^{\top} \textcolor{MatlabGreen}{D_{n}}^{\top} D_{12,j}^{\top} \right) \textcolor{MatlabBlue}{\mathcal{Z}} \\ & \quad  -  \textcolor{MatlabBlue}{\mathcal{Z}}^{\top}  \left( C_{1,j} + D_{12,j} \textcolor{MatlabGreen}{D_{n}} C_{2} \right),  \end{split}\\
\begin{split}
N_{23} &= \textcolor{MatlabGreen}{X_{1}} B_{1,j} + \textcolor{MatlabGreen}{B_{n}} D_{21,j} \\ & \quad  - \textcolor{MatlabBlue}{\mathcal{Z}}^{\top} \left( D_{11,j} + D_{12,j} \textcolor{MatlabGreen}{D_{n}} D_{21,j} \right) \\ & \quad  - \left(  C_{1,j}^{\top} + C_{2}^{\top} \textcolor{MatlabGreen}{D_{n}}^{\top} D_{12,j}^{\top} \right) \textcolor{MatlabBlue}{\mathcal{X}}, \end{split}\\
N_{24} &= \textcolor{MatlabBlue}{\mathcal{Z}}^{\top}, \\
N_{33} &= \textcolor{MatlabRed}{\nu^{2}} I - \textcolor{MatlabBlue}{\mathcal{X}}^{\top} \left( D_{11} + D_{12} \textcolor{MatlabGreen}{D_{n}} D_{21} \right) \\ & \quad  - \left( D_{11,j}^{\top} + D_{21,j}^{\top} \textcolor{MatlabGreen}{D_{n}}^{\top} D_{12,j}^{\top} \right) \textcolor{MatlabBlue}{\mathcal{X}}, \\
N_{34} &= \textcolor{MatlabBlue}{\mathcal{X}}^{\top} , \\
N_{44} &= -I.
\end{align}
\end{subequations}

\end{document}